\documentclass[]{aa} 
\usepackage{graphicx}
\usepackage[figuresright]{rotating}
\usepackage{txfonts}
\usepackage{natbib}
\bibpunct{(}{)}{;}{a}{}{,}
\def\s-1{\rm {s^{-1}}}

\usepackage{url}

\begin{document}

   \title{Galactic interstellar turbulence across the southern sky seen through spatial gradients of the polarization vector.}
   \authorrunning{Iacobelli et al.}
   \titlerunning{Interstellar turbulence in the southern sky.} 

   \author{M. Iacobelli\inst{1,}\inst{2},
          B. Burkhart\inst{4},
          M. Haverkorn\inst{3,}\inst{1},
          A. Lazarian\inst{4},
          E. Carretti\inst{5},
          L. Staveley-Smith\inst{6,7},
          B.M. Gaensler\inst{8},
          G. Bernardi\inst{9,10,11},
          M.J. Kesteven\inst{5},
          S. Poppi\inst{12}
			}
    \offprints{M. Iacobelli}

   \institute{Leiden Observatory, Leiden University, PO Box 9513, 2300 RA Leiden, the Netherlands\\
              \email{iacobelli@strw.leidenuniv.nl}
	\and
	ASTRON, the Netherlands Institute for Radio Astronomy, Postbus 2, 7990AA, Dwingeloo, The Netherlands
	\and
        Radboud University Nijmegen, Heijendaalseweg 135, 6525 AJ Nijmegen, the Netherlands
	\and
        Astronomy Department, University of Wisconsin, Madison, 475 N. Charter St., WI 53711, USA
    \and
        CSIRO Astronomy and Space Science, PO Box 76, Epping, NSW 1710, Australia
    \and
        International Centre for Radio Astronomy Research, M468, University of Western Australia, Crawley, WA 6009, Australia  
    \and
        CAASTRO: The ARC Centre of Excellence for All-sky Astrophysics
    \and
        Sydney Institute for Astronomy, School of Physics A29, The University of Sydney, NSW 2006, Australia
    \and
        SKA SA, 3rd Floor, The Park, Park Road, Pinelands 7405, South Africa
    \and
        Department of Physics and Electronics, Rhodes University, PO Box 94, Grahamstown 6140, South Africa
    \and
        Harvard–Smithsonian Center for Astrophysics, 60 Garden Street, Cambridge, MA, 02138, USA
    \and
        INAF Osservatorio Astronomico di Cagliari, Via della Scienza I-09047 Selargius (CA), Italy }
        
   \date{Received Nov 5, 2013; accepted Mar 26, 2014}
 
  \abstract
  {}  
  {Radio synchrotron polarization maps of the Galaxy can be used to infer the  properties of interstellar turbulence in the diffuse magneto-ionic medium (MIM). In this paper, we investigate the normalized spatial gradient of linearly polarized synchrotron emission ($|\nabla \textbf{P}|/|\textbf{P}|$) as a tracer of turbulence, the relationship of the gradient to the sonic Mach number of the MIM, and changes in morphology of the gradient as a function of Galactic position in the southern sky.}
   {We used data from the S-band Polarization All Sky Survey (S-PASS) to image the normalized spatial gradient of the linearly polarized synchrotron emission ($|\nabla \textbf{P}|/|\textbf{P}|$) in the entire southern sky at $2.3$~GHz. The spatial gradient of linear polarization reveals rapid changes in the density and magnetic fluctuations in the MIM due to magnetic turbulence as a function of Galactic position. We made comparisons of these data to ideal MHD numerical simulations. To constrain the sonic Mach number ($M_{s}$), we applied a high-order moments analysis to the observations and to the simulated diffuse, isothermal ISM with ideal magneto-hydrodynamic turbulence.}
  {We find that polarization gradient maps reveal elongated structures, which we associate with turbulence in the MIM. Our analysis indicates that turbulent MIM is in a generally transonic regime. This result for the turbulent regime is more general than the ones deduced by the analysis of electron density variation data, because it is based on the stochastic imprints of the Faraday rotation effect, which is also sensitive to the magnetic field fluctuations. Filamentary structures are seen with typical widths down to the angular resolution, and the observed morphologies closely match  numerical simulations and, in some cases, H$\alpha$ contours. The $|\nabla \textbf{P}|/|\textbf{P}|$ intensity is found to be approximately log-normal distributed. No systematic variations in the sonic Mach number are observed as a function of Galactic coordinates, which is consistent with turbulence in the WIM, as inferred by the analysis of H$\alpha$ data. We conclude that the sonic Mach number of the diffuse MIM appears to be spatially uniform towards the Galactic plane and the Sagittarius-Carina arm, but local variations induced by nearby extended objects are also found.}
   {}

\maketitle

\section{Introduction}
\label{intro}
Galactic magnetic fields and matter (i.e. atoms, ions and molecules) spread between the stars constitute a complex dynamic plasma, known asthe interstellar medium (ISM). The density and temperature of the particles and the magnitude of the fields are fundamental parameters that shape the structure of the interstellar environment and characterize its evolution. 
Earlier studies \citep[for a review see e.g.][]{Ferriere01,Cox05} have pointed out the presence of magnetohydrodynamic (MHD) turbulence in the ISM, which is responsible for the distribution and the dissipation of energy through a wide range of spatial scales. MHD turbulence is thought to play an essential role for many key interstellar processes \citep[for a review see e.g.][]{Elmegreen04}, including star formation \citep[see][]{KrumholzMcKee05,McKeeOstriker07}, cosmic ray propagation \citep[see][]{Schlickeiser11,Lazarian11} and magnetic reconnection \citep[see][]{LazarianVishniac99}. Additionally, astrophysical MHD turbulence is an integral part of the dynamics of the Galaxy, providing a significant pressure (and energy density) to support the diffuse ISM \citep{Boulares90}.

Many efforts have been made over the past decades to characterize magnetic fields and turbulence in the ISM as well as their mutual dependence \citep[see review by][]{BurkhartLazarian12a}. The presence of a turbulent cascade in the ISM was obtained by tracing density variations in the warm ionized medium \citep{Armstrong95,ChepurnovLazarian10}. However, MHD turbulence is traced in the different phases of the ISM by several typical signatures, such as density, velocity \citep{Pogosyan09,Chepurnov10}, and synchrotron intensity \citep{LazarianPogosyan12,Iacobelli13} variations.
Because observations of astrophysical MHD turbulence and magnetic fields are challenging, the fundamental parameters of ISM turbulence, such as the sonic and Alfv\'{e}nic Mach numbers, the magnetic field structure and strength, the Prandtl and Reynolds numbers, and the physical scale of energy injection are still poorly constrained. Therefore observational studies of MHD turbulence in the ISM, combined with analytic predictions and numerical simulations, are essential. 
Radio observations are a fundamental tool for gaining insight into magnetic fields, the density of the ionized gas, and their turbulent fluctuations. In particular, radio polarization maps constitute a useful diagnostic for studying turbulence and magnetic fields in the diffuse, ionized ISM \citep[see e.g.][]{Wieringa93,Gaensler01,Haverkorn04a,Haverkorn04b,Schnitzeler07}. Surveys covering a large part of the sky add information on the spatial dependence of these fields as well. Previous surveys of the southern sky are affected by several limitations such as incomplete sky coverage \citep[e.g. the survey at 2.4~GHz by][]{Duncan95,Duncan97}, the absence of polarimetric data \citep[e.g. the survey at 2.3~GHz by][]{Jonas85,Jonas98} as well as a limited angular resolution \citep[e.g. the survey at 1.4~GHz by][]{Testori08}. The S-band Polarization All Sky Survey (S-PASS) \citep{Carretti13} is a recent spectro-polarimetric survey of the entire southern sky carried out with the Parkes 64m telescope at 2.3~GHz to diminish depolarization effects with respect to 1.4~GHz surveys.
The use of the spatial gradient of the polarization vector to image the small-scale structure associated with ISM turbulence has been recently discovered by \citet{Gaensler11} and exploited by \citet{Burkhart12}. They show how to map the magnetized turbulence in diffuse ionized gas from the gradient of the Stokes~Q and U pseudo-vectors. In a 18~deg$^{2}$ patch of the Galactic plane, they find an intricate filamentary network of discontinuities in gas density and magnetic field. In agreement with the result of \citep{Hill08} for the warm ionized medium, these authors find turbulence in the magneto-ionic medium (MIM) to be transonic, with a sonic Mach number $M_{s}\lesssim2$ and therefore weakly compressible. These results were partially derived from the ability of statistical moments to characterize the sonic Mach numbers from spatial gradient maps of linear polarization by testing the sensitivity of statistical methods to different regimes of turbulence.  Analyses of \citet{Gaensler11} and \citet{Burkhart12} find correlations between the spatial morphology, the higher order moments of the distribution of the spatial gradients of polarized emission and the sonic and Alfv\'{e}nic Mach number.

In this paper we present the first mapping of different regimes of turbulence in the diffuse, ionized ISM over the entire southern sky by applying the statistical moments analysis to several regions. In particular, by comparing the analysis of high order moments in both simulations of MHD turbulence and observations we search for spatial variations of the sonic Mach number. Galactic coordinates are used throughout the following sections. 

In Sect.~\ref{s:obs} we present an overview of both the data and the gradient method. In Sect.~\ref{s:imaging} we present the spatial gradient map of the polarization vector displaying an extended network of filaments, and in Sect.~\ref{s:regimes_turbulence} different regimes of MHD turbulence in the ISM are characterized from the moment map analysis. Finally, we discuss our results and present conclusions in Sect.~\ref{s:conclusion}.


\section{Data overview}
\label{s:obs}

The S-band Polarization All Sky Survey (S-PASS) is a single-dish polarimetric survey of the entire southern sky at 2.3~GHz, performed with the Parkes 64~m Radio Telescope and its S-band Galileo receiver, which is a circular polarization system suitable for Stokes~Q and U measurements. S-PASS observational parameters are given in Table~\ref{t:data_prop}, while a description of S-PASS observations and analysis is given by \citet{Carretti13} and Carretti et al. (in preparation). To realize absolute polarization calibration of the data, an innovative observing strategy based on long scans along the horizon towards the east and west of the Parkes telescope was adopted. A system temperature of about 20~K was reached, with an improvement of a factor of 2 with respect to the previous continuum survey at the same frequency \citep[see e.g.][]{Duncan97}. Another main feature of the survey is its high angular resolution; due to a telescope beam width of $8.9\arcmin$, final maps were obtained with a beam of $FWHM=10.75\arcmin$ with an improvement of about a factor of 4 with respect to a previous continuum all-sky survey at $1.4$~GHz \citep[see e.g.][]{Reich01}. The final maps of the Stokes parameters resulting from the S-PASS survey will be fully presented by Carretti et al. (2013, in preparation). We used the Stokes~Q and U maps to image the spatial gradients of the linearly polarized emission for our analysis of ISM turbulence.

\begin{table}
\centering      
\tiny
\caption{\label{t:data_prop} Observational properties of S-PASS data.}
\begin{tabular}{ll}
 & \\
\hline
\hline \\ 
Reference frequency & $\nu = 2307$~MHz \\
Bandwidth & $\Delta\nu = 184$~MHz \\
Gridded beamwidth & $FWHM = 10.75\arcmin$ \\
Stokes~Q,U maps rms & $\sigma \lesssim 1$~mJy~beam$^{-1}$ \\
Gain (Jy/K) at 2307~MHz & 1~mJy = 0.55~mK \\
\\
\hline \\
\end{tabular} 

\end{table}

\subsection{Radio polarization gradients}
\label{s:pol_grad}

Variations in polarized intensity and polarization angles of radio synchrotron radiation are related to magnetic turbulence in the ISM. However, both these quantities are not invariant to rotations and translations in the Q--U plane, e.g. due to Faraday rotation by a uniform foreground screen and incomplete interferometric sampling respectively, making their interpretation difficult. A new diagnostic to investigate the turbulent fluctuations in the ISM affecting the radio polarization measurements is the spatial gradient of the polarization vector  \citep{Gaensler11}.

The radio polarization gradient ($\nabla \textbf{P}$) indicates the variation in the polarization vector (\textbf{P}) as a function of position in the image plane. In addition, the spatial gradient of the polarization vector acts as a high-pass filter, allowing one to recover interesting features from a map with a strong DC offset \citep[see e.g.][]{Schnitzeler09}. To calculate the gradient of the polarization vector, we follow \citet{Gaensler11}, determining the magnitude as:
\begin{equation}
|\nabla \textbf{P}| = \sqrt{\left(\frac{\partial Q}{\partial x}\right)^{2} + \left(\frac{\partial U}{\partial x}\right)^{2} + \left(\frac{\partial Q}{\partial y}\right)^{2} + \left(\frac{\partial U}{\partial y}\right)^{2}} \, .
\end{equation}
In general, fluctuations in $\vec{P}$ are due to fluctuations in emission, as well as fluctuations in Faraday rotation \citep{Spangler82,Spangler83}. However, in this paper we treat the fluctuations in $\vec{P}$ as the result of stochastic Faraday rotation due to Faraday screens on a polarized background. In this Faraday thin approximation, the magnitude of the gradient of the polarization vector traces the Faraday rotation changes from one line of sight to another, as modelled in \citet{Burkhart12}. We use correlation coefficients (see Section~\ref{s:regimes_turbulence}) to show that this approximation is justified for the fields of view under consideration.
Gradients in the polarization vector can be caused by fluctuations in different quantities, as detailed below.
\begin{enumerate}
\item Fluctuations in Stokes~I alone. Even in the case of no Faraday rotation or depolarization, variations in Stokes~I will cause variations in the polarized intensity ($|\textbf{P}|$) and therefore a non-zero $|\nabla \textbf{P}|$. In this case, $|\textbf{P}| \propto I$ and $|\nabla \textbf{P}| \propto \nabla I$. Maps of $|\nabla \textbf{P}|$ will act as an edge detection algorithm for fluctuations in synchrotron emission.

\item Fluctuations in polarization angle alone. This is the case of a Faraday screen, which alters the direction of the polarization vector but not its amplitude. Then, $|\textbf{P}| \propto I$, but $|\nabla \textbf{P}|$ is not correlated with either $|\textbf{P}|$ or Stokes~I. In this case, $|\nabla \textbf{P}|$ directly traces gradients in rotation measure (RM), hence turbulence in the magneto-ionic ISM. This situation can be divided into two special cases:

\begin{enumerate}
\item A Faraday screen in front of spatially constant synchrotron emission. In this case, both Stokes~I and $|\textbf{P}|$ are constant, but $|\nabla \textbf{P}|$ is not constant. Then  \citep{Burkhart12},
\begin{equation} \label{eq:eqBurkhart}
\vert \nabla RM \vert = \frac{|\nabla\textbf{P}|}{2\lambda^{2}|\textbf{P}|}
\end{equation}
and the quantity $|\nabla \textbf{P}|/|\textbf{P}|$ should be used to track fluctuations in RM, i.e. to track magnetic turbulence. This is also the case in the simulations we use to compare to the data.

\item The Faraday screen emits thermal, unpolarized radiation, such as  an \ion{H}{ii} region. In this case, some correlation between Stokes~I and $|\nabla \textbf{P}|$ may exist.
\end{enumerate}

\item Fluctuations in the polarized intensity and polarization angle, as caused by depolarization. In this case, equation~\ref{eq:eqBurkhart} does not hold. Turbulence, as traced by $|\nabla RM|$, is tracked by $|\nabla \textbf{P}|$. This is the situation for which the polarization gradient method was originally created \citep{Gaensler11}: RM fluctuations that cause depolarization canals are masked by foreground/background polarized emission, but the polarization gradient method makes these RM fluctuations visible even in the presence of a polarized emission foreground/background.
\end{enumerate}

In the observations, gradients in the polarization vector will be caused by a combination of these three effects. In the following, we estimate the importance of each of these three effects by investigating correlations among Stokes~I, polarized intensity $|\textbf{P}|$, and polarization gradient $|\nabla \textbf{P}|$. In fields where the Faraday screen approximation is reasonable, the normalized gradient map $|\nabla \textbf{P}|/|\textbf{P}|$ traces fluctuations in RM. However, in fields where this approximation does not hold, the un-normalised map $|\nabla \textbf{P}|$ should be considered.

Both the magnitude of the polarization spatial gradient ($|\nabla \textbf{P}|$) and of the polarization vector ($|\textbf{P}|$) follow an asymmetric positive distribution, while the Stokes~Q and U parameters follow Gaussian distributions. The positive definite distributions are signal-to-noise (S/N) dependent and are thus affected by a bias towards high values at low S/N \citep{Vinokur65}. \citet{Wardle74} suggested a widely used correction for the bias of a positive defined distribution $P(R)$ as 
\begin{equation} 
R \sim R_{M}^{\prime} \left[1 - \left( \frac{\sigma^{\prime}}{R_{M}^{\prime}} \right) \right]^{1/2} \, ,
\end{equation}
where $R_{M}^{\prime}$ is the measured magnitude, $\sigma^{\prime}$ the measured noise level, and \textit{R} the intrinsic magnitude of the signal vector. We confirmed the Gaussian behaviour of Stokes~Q and U maps and determined a noise level of $1$~mJy~beam$^{-1}$ from a small region assumed to contain no emission from the sky. We also assumed uniformity of the noise across the entire field. Finally, we calculated the normalized gradient of the polarization vector and  discarded lines of sight where \textbf{P} is at the level of the bias 
$|\textbf{P}_{R}| = 1.25\sigma_{Q,U}$, resulting in a fraction of blanked pixels of about 1\% in the final $|\nabla \textbf{P}|/|\textbf{P}|$ map.

To quantify the impact of the foreground Faraday rotation effects for these data, we looked at the correspondence between total intensity and polarised intensity, by determining the degree of correlation between the Stokes~I and the $|\textbf{P}|$ maps. If the polarization signal is due to foreground Faraday rotation, little linear correlation between the fluctuations of $|\textbf{P}|$ and Stokes~I is expected. Thus, we adopted the Pearson correlation coefficient ($\rho_{p}$) defined as
\begin{equation}
\rho_{p} = \frac{cov\,(P,I)}{\sigma_{P}\,\sigma_{I}} \quad,
\end{equation}
where $cov\,(P,I)$ is the covariance and $\sigma_{P}$, $\sigma_{I}$ are the standard deviations of the Stokes~I and the $|\textbf{P}|$ maps. We considered the range of weak correlation to be $|\rho_{p}|\leq0.35$ and find a mild positive correlation $\rho_{p}=0.38$, so the bulk of the $|\nabla \textbf{P}|/|\textbf{P}|$ fluctuations are not intrinsic to the sources of polarized emission and can be interpreted in the framework of MHD turbulence in the ionized ISM. In the following we select regions of low correlation where the $|\nabla \textbf{P}|/|\textbf{P}|$ fluctuations can be interpreted in the framework of MHD turbulence.



\begin{sidewaysfigure*}
\vspace{180mm}
\hspace{0mm}
\resizebox{25cm}{!}{\includegraphics[]{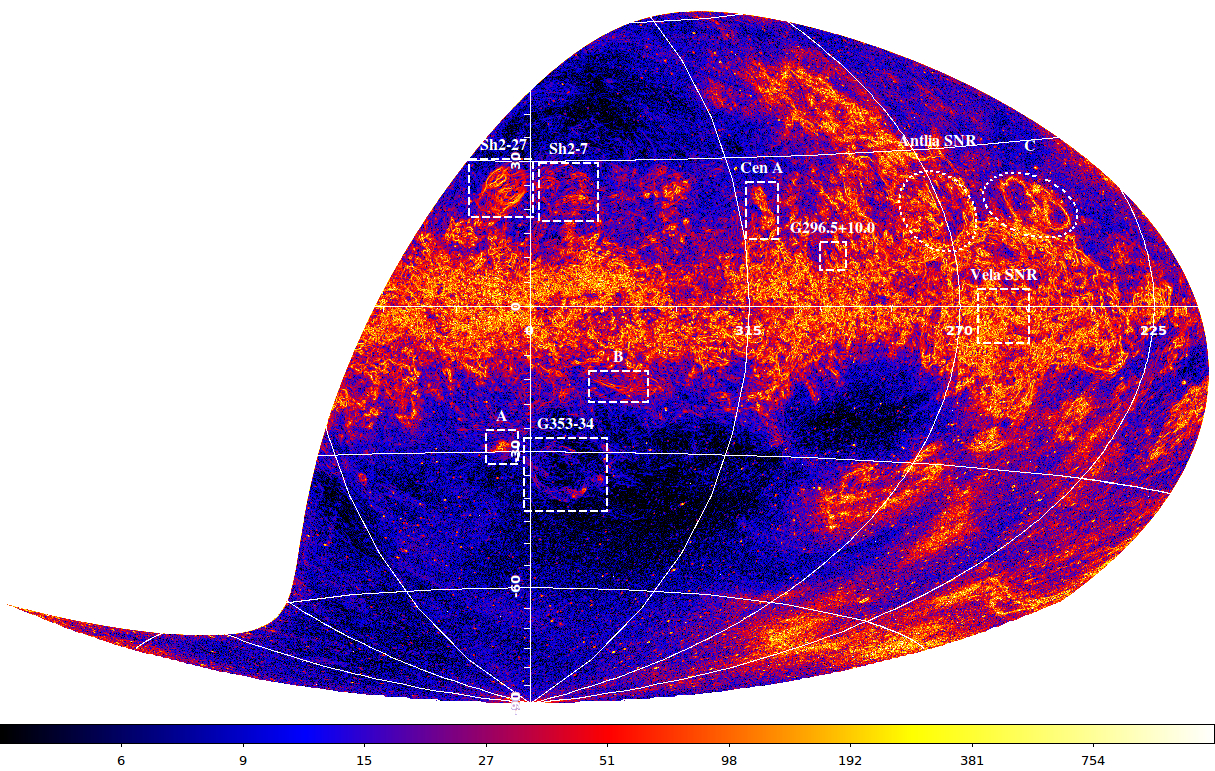}}
\caption{\label{f:gradPInorm_map} \textit{The map of the normalised spatial gradient of linearly polarized emission, $|\nabla \textbf{P}|/|\textbf{P}|$, in Aitoff projection of the entire southern sky at a reference frequency of $2.3$~GHz and a resolution of $11\arcmin$. The noise level in the map was obtained by averaging intensity over a small region with no signal and intensity is clipped at 3~$\sigma$. The scaling runs from 0.2 (black) to 1000 (white). Intensity unit is beam$^{-0.5}$.}}
\end{sidewaysfigure*}


\section{Observational results}
\label{s:imaging}

Figure~\ref{f:gradPInorm_map} shows the final de-biased $|\nabla \textbf{P}|/|\textbf{P}|$ sky map, clipped at 3~$\sigma$ for displaying purpose. 
The spatial gradient of the astrophysical polarized synchrotron emission of the diffuse Galactic foreground is depicted in this image at 2.3~GHz, producing a wealth of structures. Faint artefacts are also present around the equatorial south pole due to the scanning strategy. We now describe the main morphological features in the de-biased $|\nabla \textbf{P}|/|\textbf{P}|$ map. 

First, there are spatially compact and isolated structures, corresponding to bright extragalactic synchrotron sources. Several hundred unresolved sources (mostly quasars and radio galaxies) are detected to be smoothly distributed on the sky plane. Furthermore, a polarized extragalactic source showing a striking extended morphology is also present and marked in Fig.~\ref{f:gradPInorm_map}, the bright radio galaxy Cen~A, with a north-south extension of about $8^{\circ}$ of its structured emitting lobes.

There are also isolated Galactic structures on a range of scales. Peaks in the normalized polarization gradient signal are seen throughout the Galactic plane up to Galactic latitudes of $\vert b \vert\approx25^{\circ}$ corresponding to bright supernova remnants (SNRs) and nearby \ion{H}{ii} regions. While the brightness of the former is mainly due to their intrinsic polarization, the latter are traced out by Faraday rotation along the line of sight to the observer induced by their thermal electrons. 
Indeed nearby (distance $\sim 200$~pc) and extended \ion{H}{ii} regions are seen towards the inner Galaxy, such as Sh\,2-27 around the O-class star $\zeta$~Ophiuchi (at $l\sim8^{\circ}$, $b\sim23^{\circ}$) and Sh\,2-7 excited by the nearby star $\delta$~Sco (at $l\sim350^{\circ}$, $b\sim22^{\circ}$). 

\begin{table}
\centering      
\tiny
\caption{\label{t:SNR_prop} List of unambiguously identified SNRs in all of Stokes~I, polarized intensity, and polarization gradient S-PASS maps, by cross-checking radio data taken from the \citet{Green09} database.}
\begin{tabular}{lccc}
 & \\
\hline
\hline \\
Name & Other names & Distance$^{a}$ \\
     &             &   [kpc] \\
\hline \\
G261.9+5.5 &       & $2.9$ \\
G263.9$-$3.3 & Vela\,(XYZ) & $0.29$ \\ 
G296.5+10.0 & PKS\,1209-51/52 & $2.1$ \\
G327.6+14.6 & SN\,1006 & $2.2$ \\
\\
\hline \\
\end{tabular}

{\it a)}\,The listed distance values are taken from \citet{Milne70}, \citet{Dodson03}, \citet{Giacani00}, and \citet{Ghavamian02}, respectively. \\

\end{table}

\begin{figure}
\resizebox{1.0\columnwidth}{!}{\includegraphics{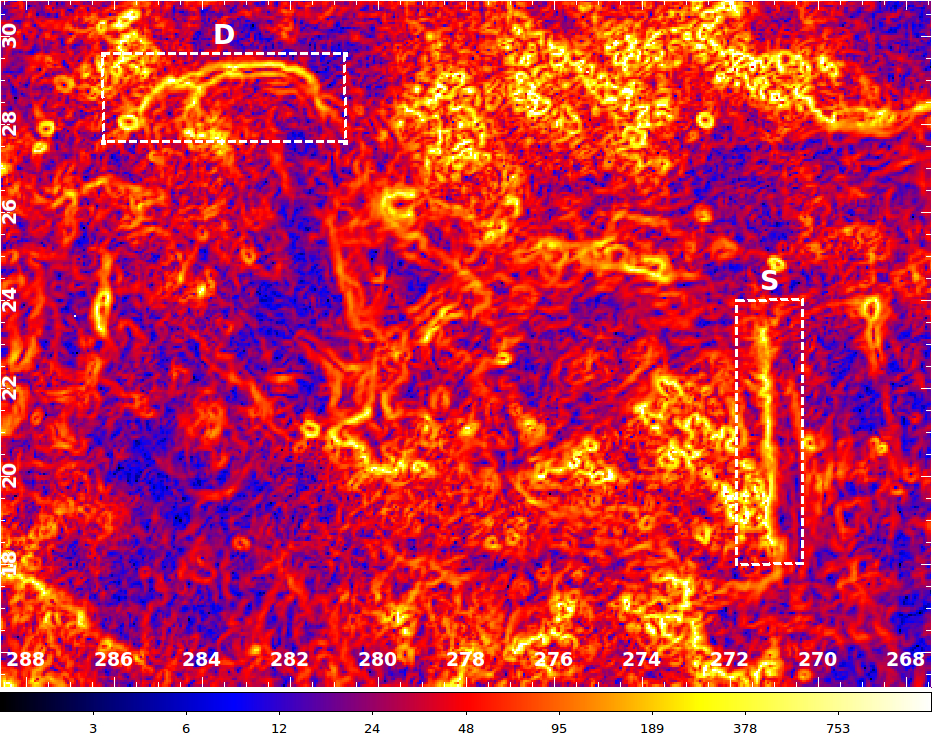}}
\caption{\label{f:gradPInorm_cases} \textit{A zoom of Fig.~\ref{f:gradPInorm_map} centred on $l\sim278^{\circ}$,$b\sim22^{\circ}$, showing different observed morphologies for different MHD turbulence cases ``double-jumps'' (D boxed region) and ``single-jump'' (S boxed region) filaments due to (strong or weak) shocks and a cusps or shears along the line of sight, respectively. Intensity scale runs from 0.2 (black) to 2000 (white).}} 
\end{figure}

The outstanding characteristic is that most of the polarization gradients come from a network of filaments, whichform clearly separated and extended regions well outside the Galactic plane. The typical width of these filamentary structures matches the angular resolution. Furthermore, different morphological patterns of these filaments are seen. In Fig.~\ref{f:gradPInorm_cases} we show an example of the observed morphology for the normalized spatial gradients of polarized synchrotron emission, as a result of fluctuations in the MIM induced by different types of discontinuities \citep{Burkhart12}. The panel depicts a complex region above the Galactic plane centred at $(l,b)\approx(278^{\circ},+22^{\circ})$ and displays filamentary structures over a wide range of angular scales, up to about six degrees, with no correspondence to Stokes~I. This extended pattern of filaments is due to the presence of discontinuities in the interstellar gas density and magnetic field. In particular, we note the presence of two evident structures that are also seen in MHD simulations \citep[for a comparison see Fig.~4 of][]{Burkhart12} and are interpreted as due to different kinds of discontinuities in the warm, ionized ISM: a ``double-jump'' profile (box D) and a ``single-jump'' profile (box S), corresponding to the case of a (strong or weak) shock and a cusp or shear along the line of sight, respectively. Moreover changes in the $|\nabla \textbf{P}|/|\textbf{P}|$ brightness along the filaments due to variations of the magnitude of turbulent fluctuations along the line of sight are recognized.

Complex large-scale structures (such as the nearby and shell-like G\,353-34) are seen up to intermediate ($\vert b\vert\approx40^{\circ}$) Galactic latitudes. A prominent arc-like feature is observed within the longitude range $270^{\circ}\lesssim l \lesssim290^{\circ}$. The base of this magnetic structure is around $l=270^{\circ}$, and it extends up as an arc from the Galactic plane up to about $(l,b)=(285^{\circ},45^{\circ})$ with an overall length of $\sim50^{\circ}$. Up to this point the structure is traced by the high polarization gradient intensity exhibiting a complex morphology: mostly filamentary at Galactic latitudes below $20^{\circ}$ and mainly patchy, small-scale rich towards higher latitudes. 
High intensity values of radio polarization gradients are seen towards the third Galactic quadrant at high and negative Galactic latitudes, forming a very extended and patchy pattern up to $b=-70^{\circ}$ which is not spatially coincident with the nearby ($d\lesssim500$~pc) region of higher interstellar polarization pointed out by \citet{Berdyugin04}. 

\subsection{The polarization horizon}

To constrain the impact of depolarization, to perform a proper comparison of data with numerical simulations, and to convert angular scales to spatial scales, we need to estimate the distance to which polarized emission is detected. 

This spatial depth is referred to as the ``polarization horizon'', and it has been used in previous works \citep[see e.g.][]{Landecker02,Uyaniker03}. This quantity is a function of the instrumental features (i.e. the beam size, the observing frequency), as well as of the physical conditions of the probed medium (causing an intrinsic degree of depolarization). The smaller the beam and/or the higher the observing frequency, the farther the polarization horizon; the brighter and/or more coherent the synchrotron emission, the farther the corresponding polarization horizon. Recently, \citet{Carretti13} have derived an estimate of about $2-3$~kpc for the ``polarization horizon'' of these S-PASS data from the analysis of depolarized regions at low latitudes towards the inner Galaxy. However, the polarization horizon depends on the viewing direction in the Milky Way, and we consider the all-sky $|\nabla \textbf{P}|/|\textbf{P}|$ intensity, so we need to find estimates for the maximum spatial depth up to intermediate Galactic latitudes ($\vert b \vert\lesssim 30^{\circ}$) and towards the outer regions ($260^{\circ} \lesssim l \lesssim 330^{\circ}$). To estimate a lower limit for the polarization horizon, we cross-checked the radio catalogue\footnote{The catalogue is available at http://www.mrao.cam.ac.uk/surveys/snrs/} of Galactic supernova remnants \citep{Green09} with detections in the Stokes~I and $|\nabla \textbf{P}|$ S-PASS maps, and we considered SNRs with a minimum angular size matching the S-PASS angular resolution. We used the $|\nabla \textbf{P}|$ map because for sources with a large intrinsic polarization component, it acts as an edge detection algorithm, making the source detection towards regions with strong diffuse emission easier. Also, we note that the non-detection of polarization from a SNR may be due to internal depolarization in the SNR itself, constituting a limitation of the adopted method. Thus we decided not to use non-detections to constrain the upper limit of the polarization horizon. 

We find four SNRs that satisfy our criteria (see Table~\ref{t:SNR_prop}) and suggest a lower limit for the polarization horizon about 3~kpc away from the Galactic centre, which agrees with the above estimate of \citet{Carretti13} and is used throughout the following sections. At this distance, the presence of radio polarization gradient features down to an angular resolution of $\sim11\arcmin$ corresponds to a linear scale of $<10$~pc for the filament widths, and therefore for the MIM fluctuations in the disk and the halo traced by these filaments. The earlier detection of $|\nabla \textbf{P}|/|\textbf{P}|$ filaments at arcmin resolution by \citet{Gaensler11}, however, implies a typical width of $\approx 1$~pc, if a polarization horizon of about 3~kpc is assumed for their data.\\

\subsection{Galaxy-scale variations in $|\nabla \textbf{P}|/|\textbf{P}|$}
\label{s:variations}

The above lower limit of about 3~kpc for the polarization horizon constrains our data to probe the nearby, major large-scale Galactic features: the Local arm (Orion Spur) towards the outer region (i.e. $210^{\circ} \lesssim l \lesssim 280^{\circ}$) and the Carina-Sagittarius arm located at a distance of about 2~kpc \citep{Georgelin00} between the Centaurus-Scutum spiral arm ($\sim3.5$~kpc) and the Sun towards the fourth Galactic quadrant. Also some emission from the nearest side of the Centaurus-Scutum spiral arm towards the Galactic center and the Perseus arm towards the outer region (i.e.\ $l \lesssim 250^{\circ}$) cannot be excluded. Moreover, out of the Galactic disk our estimate of about 3~kpc for the polarization horizon indicates a negligible amount of depolarization. Indeed, at Galactic latitudes $\vert b \vert \gtrsim 45^{\circ}$ this path length corresponds to a height $\gtrsim 2.1$~kpc, which is consistent with the scale heights of the relevant ISM components, the synchrotron-emitting thick disk, \citep[][]{Beuermann85,Kobayashi04} and the scale height of the free electron density \citep{Gaensler08,Schnitzeler12}.

In this data set, regions of high $|\nabla \textbf{P}|/|\textbf{P}|$ with no Stokes~I correlation effectively trace the loci of compression or rarefactions of the magnetic fields and/or the warm ionized gas in the ISM, as well as abrupt changes in the direction of the magnetic field. Because the spiral arms constitute the most extended and synchrotron-emitting Galactic structures where these compressions and rarefactions take place, a spatial dependence of the $|\nabla \textbf{P}|/|\textbf{P}|$ intensity with Galactic latitude and longitude is expected. 

\begin{figure}
\resizebox{9cm}{!}{\includegraphics[angle=0]{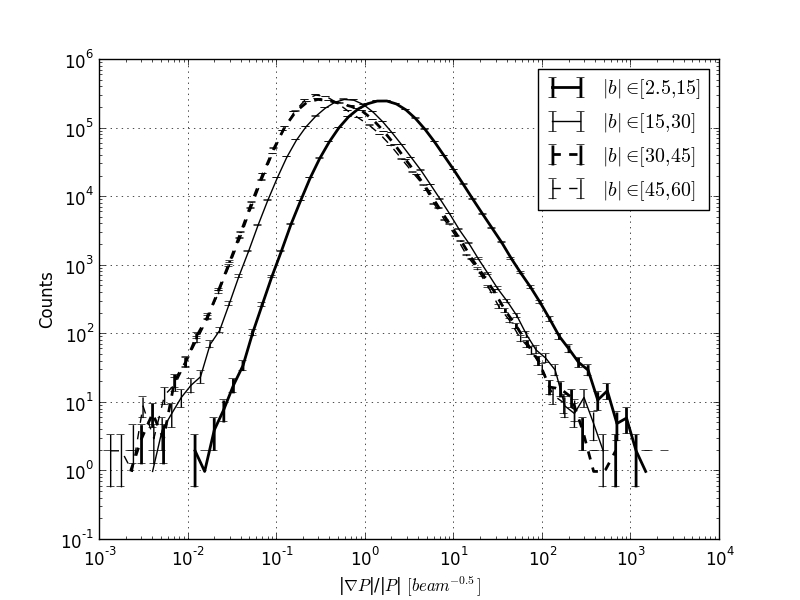}}
\caption{\label{f:Gb_counts_gradpi} \textit{Distributions of the $|\nabla \textbf{P}|/|\textbf{P}|$ intensity of the masked map at four ranges of Galactic latitude ($|b|$). Statistical errors in each intensity bin are also shown.}}
\end{figure}

To focus on the diffuse emission, we mask the Galactic plane ($|b|\leq 2.5^{\circ}$) where depolarization effects are strong. The resulting map displays a diffuse foreground and a background consisting of extragalactic discrete sources and noise. We explore the dependence on Galactic latitude by looking at the distribution of the $|\nabla \textbf{P}|/|\textbf{P}|$ intensity over four latitude ranges (see Fig.~\ref{f:Gb_counts_gradpi}). The distribution peak position of the $|\nabla \textbf{P}|/|\textbf{P}|$ intensities decreases with distance from the Galactic plane, but not the distribution width. In addition, the intensity counts are approximately represented by a log-normal distribution and the log-normal fit to the $|\nabla \textbf{P}|/|\textbf{P}|$ intensity distribution of the latitude bin $|b|\lesssim15^{\circ}$, which mostly probes the diffuse foreground emission, shows a reduced-$\chi^{2}$ of 2.09. Increasing deviations from a log-normal distribution are found with Galactic latitude.


\begin{figure}
\resizebox{9cm}{!}{\includegraphics[angle=0]{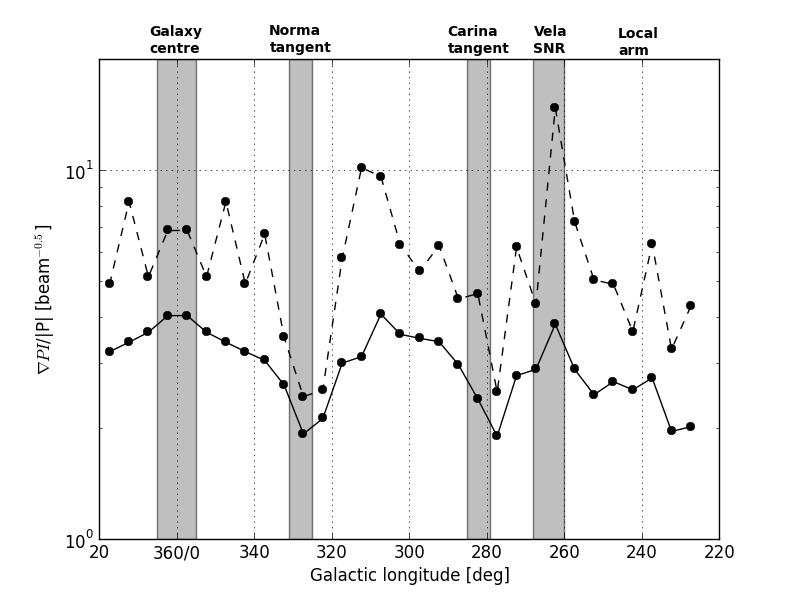}}
\caption{\label{f:mom_gradpi} \textit{The mean (solid lines) and standard deviation (dashed lines) profiles towards the Galactic disk ($5^{\circ} \lesssim \vert b \vert \lesssim 10^{\circ}$) for the $|\nabla \textbf{P}|/|\textbf{P}|$ intensity as a function of Galactic longitude. The ranges of Galactic longitudes corresponding to the main Galactic features and spiral arms are also indicated. Data points are obtained from square regions of size  $5^{\circ}$.}}
\end{figure}

We highlight the presence of a longitude dependence of the $|\nabla \textbf{P}|/|\textbf{P}|$ emissivity by using the first two moments (i.e. the mean and standard deviation) of the $|\nabla \textbf{P}|/|\textbf{P}|$ distribution. We select lines of sight out of the Galactic disk over the latitude range $5^{\circ} \lesssim \vert b \vert \lesssim 10^{\circ}$ to limit the impact of depolarization. We sample the range of Galactic longitude with $5^{\circ}$ sized squared bins, from which the mean and standard deviation of the $|\nabla \textbf{P}|/|\textbf{P}|$ intensity are calculated and displayed in Fig.~\ref{f:mom_gradpi}. We use the mean and standard deviation of $|\nabla \textbf{P}|/|\textbf{P}|$ map as tracers of the diffuse emission and the presence of compact sources (e.g. SNRs, \ion{H}{ii} regions), respectively. Both the profiles are characterized by the presence of peaks corresponding to nearby extended structures (e.g. the Vela SNR), as well as large-scale Galactic structures (e.g. the Carina-Sagittarius spiral arm), and a longitude dependence is seen with mean value of the normalized polarization gradient gradually decreasing towards the outer region. Towards the anti-centre (i.e. $l\lesssim 240^{\circ}$), no relevant diffuse emission is detected and few compact sources are seen. These directions mostly probe the Local-arm synchrotron diffuse emission, so both the $|\nabla \textbf{P}|/|\textbf{P}|$ mean and standard deviation have low values, perhaps due to depolarization of the far end of the arm. A sudden jump in both moments is found around $l\sim 260^{\circ}$, corresponding to the extended and nearby Gum Nebula and Vela SNR complex. Two minima are seen at $l_{MIN,0}=(277.5^{\circ}\pm2.5^{\circ})$ and $l_{MIN,1}=(327.5^{\circ}\pm2.5^{\circ})$, corresponding to the observed tangential directions \citep[see Table~2 of][]{Vallee08} of the Carina ($282^{\circ}\pm2^{\circ}$) and Norma ($328^{\circ}\pm3^{\circ}$) spiral arms. The low values of the mean and standard deviation may be explained by depolarization of the far end of the arms or by the alignment between the line of sight and the direction of the large-scale magnetic field. The latter option would imply a drop in the synchrotron emissivity.


\subsection{Thermal electron density and magnetic features}

The presence of Faraday rotation effects in these data are evident when comparing the $|\nabla \textbf{P}|/|\textbf{P}|$ map with the H$\alpha$ map. In this way it is possible to highlight (nearby) extended structures with relevant density and/or magnetic fluctuations. To this aim, we use the all-sky H$\alpha$ map of \citet{Finkbeiner03}, which has an angular resolution (of 6\arcmin) similar to that of the $|\nabla \textbf{P}|/|\textbf{P}|$ map. We now report on some structures induced by density and/or magnetic fluctuations.

\subsubsection{Nearby \ion{H}{ii} regions}

\begin{figure}
\resizebox{9cm}{!}{\includegraphics[angle=0]{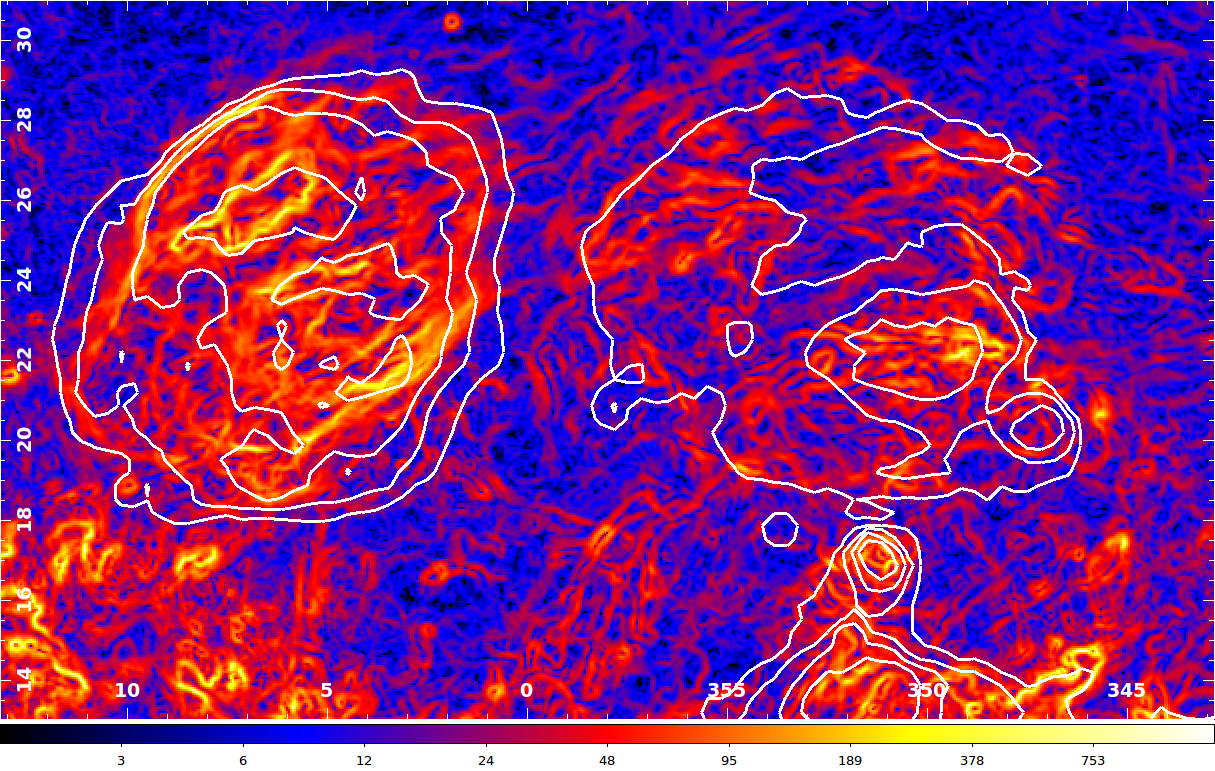}}
\caption{\label{f:Hii_maps} \textit{The bubble-like $|\nabla \textbf{P}|/|\textbf{P}|$ morphology of the nearby \ion{H}{ii} regions Sh~2-27 (on the left) and Sh~2-7 (on the right) is shown together with the contours of the H$\alpha$ intensity from the all-sky map of \citet{Finkbeiner03}. The units of contours are 1, 13, 25, 50 and 100 Rayleigh. Colour scale is the same as in Fig.~\ref{f:gradPInorm_cases}.}}
\end{figure}

Nearby \ion{H}{ii} regions are extended objects with prominent electron density fluctuations, for which a correspondence with H$\alpha$ intensity is expected. Indeed a clear correlation of the $|\nabla \textbf{P}|/\textbf{P}|$ and H$\alpha$ intensity patterns is seen towards the nearby \ion{H}{ii} regions Sh~2-27 and Sh~2-7, as shown in Fig.~\ref{f:Hii_maps}. The H$\alpha$ intensity contours clearly trace features with higher $|\nabla \textbf{P}|/|\textbf{P}|$ intensity both at the edges and towards the centre. A weak radio continuum feature is only seen for Sh~2-27, and the degree of correlation between the $|\nabla \textbf{P}|/\textbf{P}|$ and the Stokes~I intensity is $|\rho_{p}|=0.30,0.53$ for Sh~2-27 and Sh~2-7, respectively. The higher degree of correlation found towards Sh~2-7 can be explained by the overlapping with the Galactic central spur, a feature detected with S-PASS in Stokes~I and polarization \citep{Carretti13}, and possibly emanating from the Galactic centre. The interpretation of the observed morphology in terms of ISM turbulent fluctuations is therefore supported only for Sh~2-27. This bubble-like feature is crossed and filled by filaments exhibiting a ``single-jump'' morphology, itself suggesting the presence of weak shock and/or strong turbulent fluctuations \citep{Burkhart12}.


\subsubsection{Nearby, old SNRs}

\begin{figure}
\resizebox{9cm}{!}{\includegraphics[angle=0]{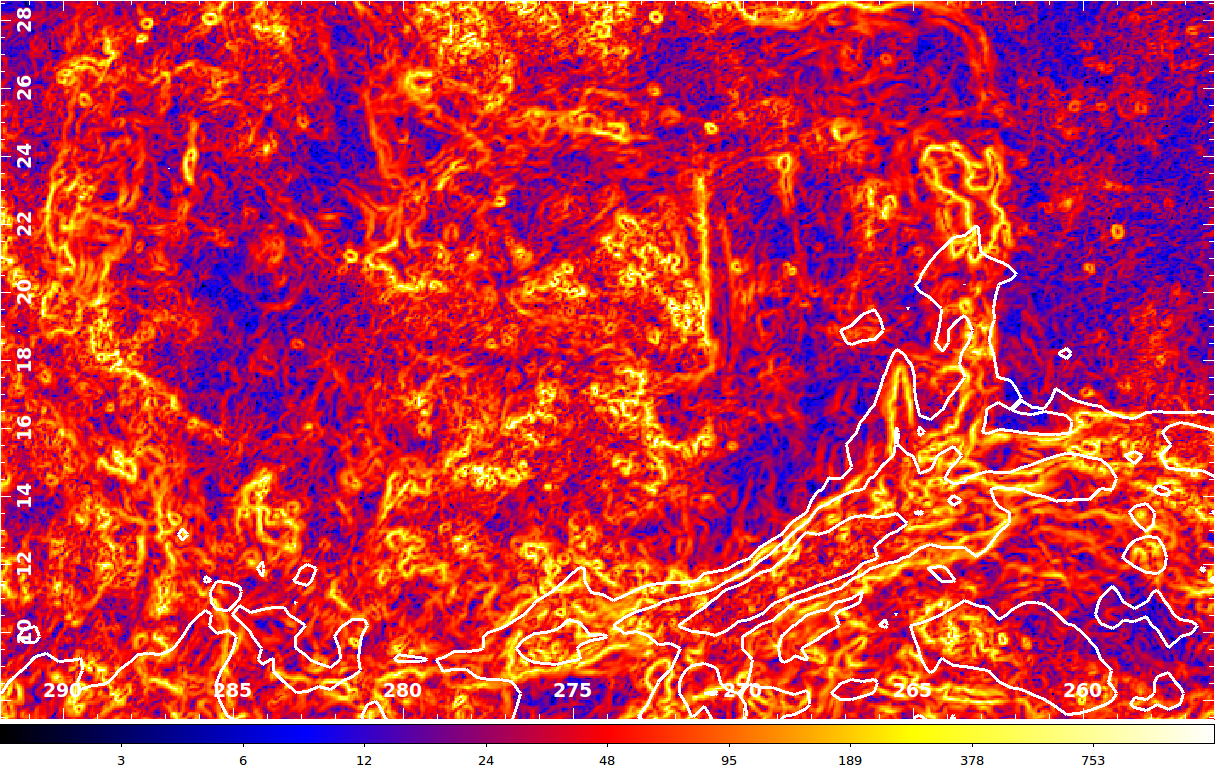}}
\resizebox{9cm}{!}{\includegraphics[angle=0]{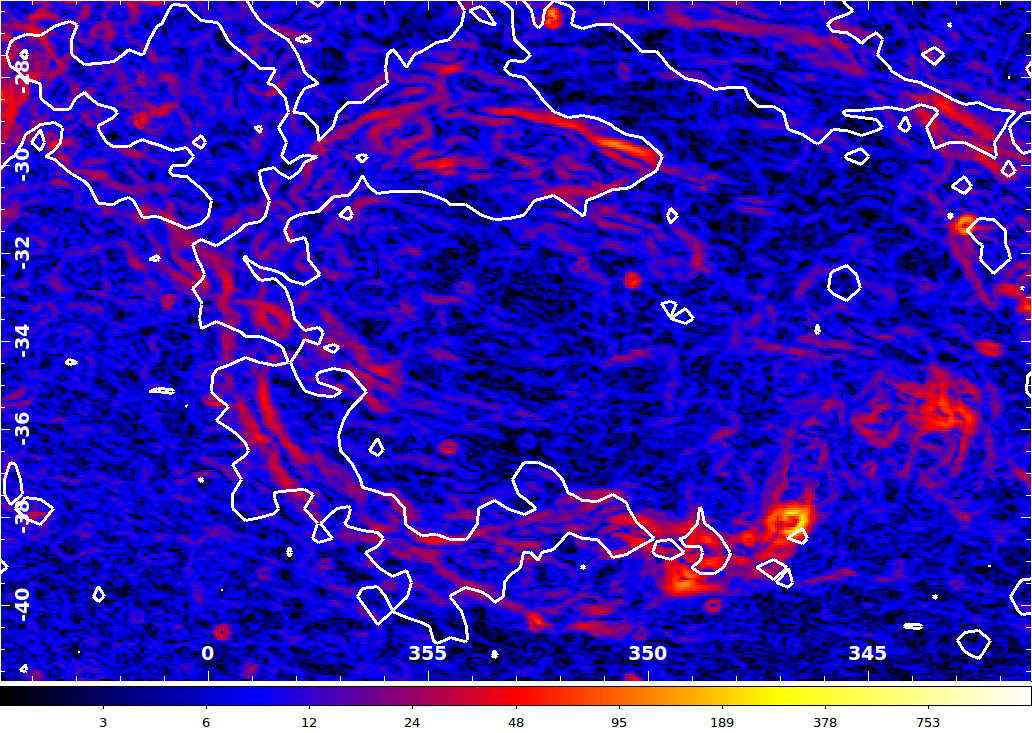}}
\caption{\label{f:shell_maps} \textit{The $|\nabla \textbf{P}|/|\textbf{P}|$ morphologies of the old SNRs Antlia (top) and G\,353-34 (bottom) are shown together with the contours of the H$\alpha$ intensity from the all-sky map of \citet{Finkbeiner03}. The units of contours for Antlia and G\,353-34 are 0.1, 20, 40, 80, 160 and 0.1, 1.5, 8, 16, 32 Rayleigh, respectively. Intensity scale runs from 0.2 (black) to 2000 (white) and from 0.2 (black) to 1000 (white) for Antlia and G\,353-34, respectively.}}
\end{figure}

We find an evident correlation of $|\nabla \textbf{P}|/\textbf{P}|$ and H$\alpha$ intensity patterns for two extended shells very likely caused by nearby ($d_{SNR}\lesssim500$~pc), old ($t_{SNR}\approx1$~Myr) SNRs: the Antlia \citep{McCullough02} and G\,353-34 \citep{Testori08} SNRs. 
The radio features are seen as incomplete shells around the centre and partially filled by fainter gradients in polarized emission. Moreover, the Antlia SNR feature is seen in a field with a complex pattern of $|\nabla \textbf{P}|/|\textbf{P}|$ intensity with no H$\alpha$ correspondence. The H$\alpha$ intensity contours show prominent arc features throughout the old radio shell. 
Unlike the younger SNRs listed in Table~\ref{t:SNR_prop}, both these evolved objects show prominent H$\alpha$ and weak radio continuum features, and an impressive morphology of filaments in polarization gradients is seen matching the H$\alpha$ features (see Fig.~\ref{f:shell_maps}). These extended and nearby structures in polarization gradients and H$\alpha$ are surrounded by and likely interacting with complex environment in the ISM, as suggested by the depolarization seen towards the brightest H$\alpha$ features along the two shells. 


\subsubsection{$|\nabla RM|$ features A,B,C}

\begin{figure}[h!tbp]
\resizebox{9cm}{!}{\includegraphics[angle=0]{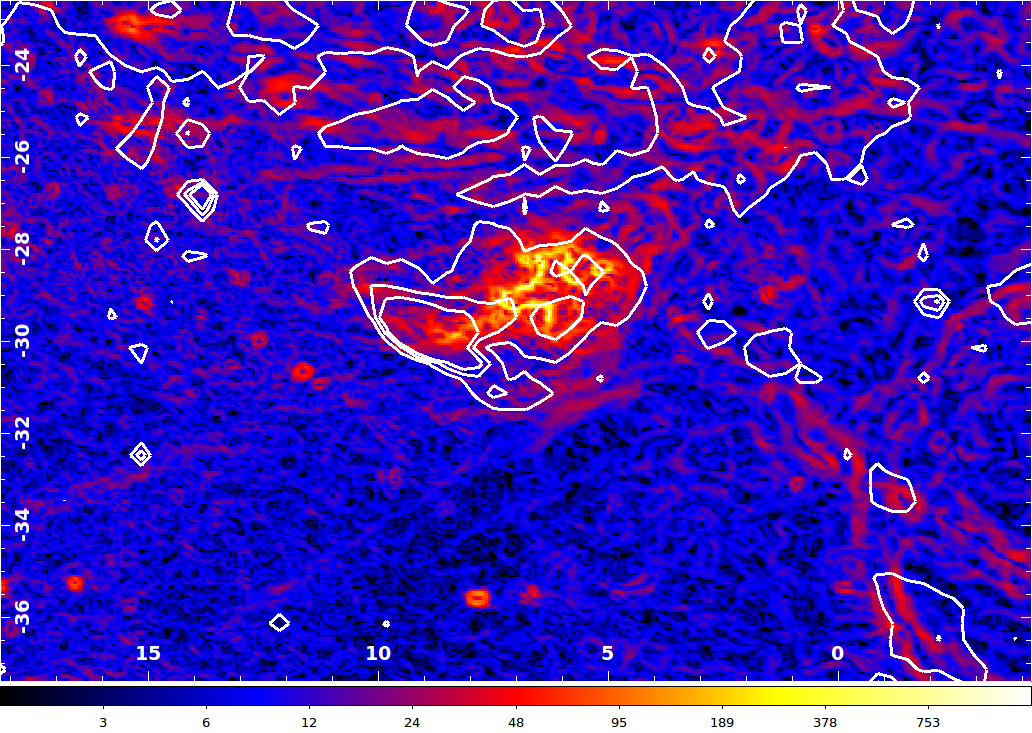}}
\resizebox{9cm}{!}{\includegraphics[angle=0]{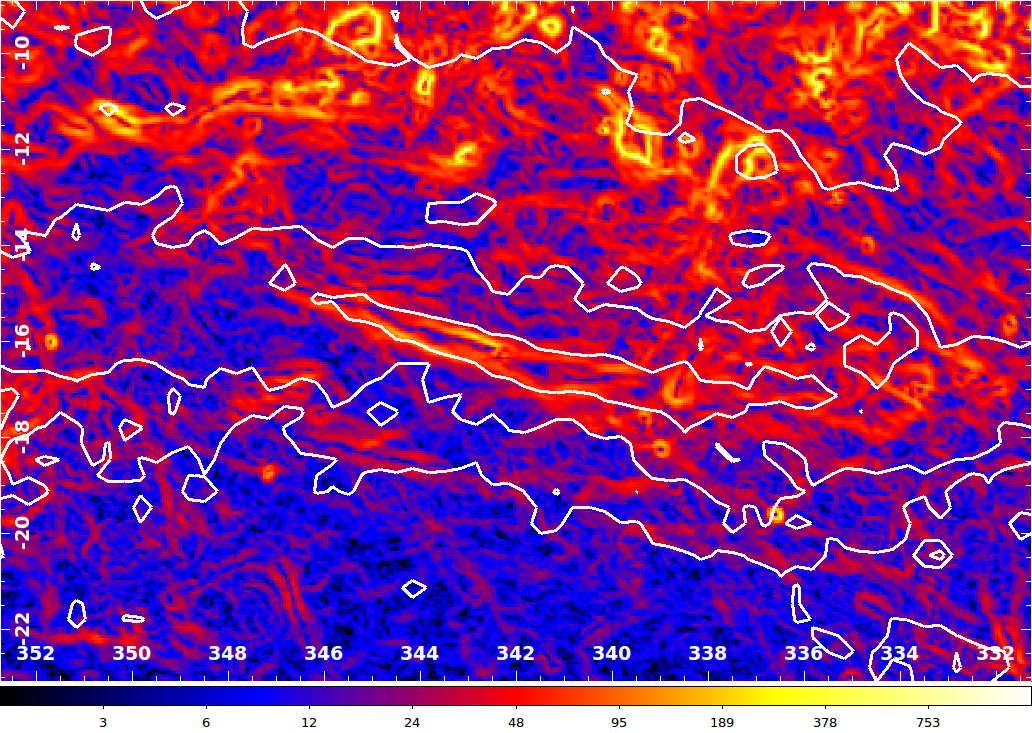}}
\resizebox{9cm}{!}{\includegraphics[angle=0]{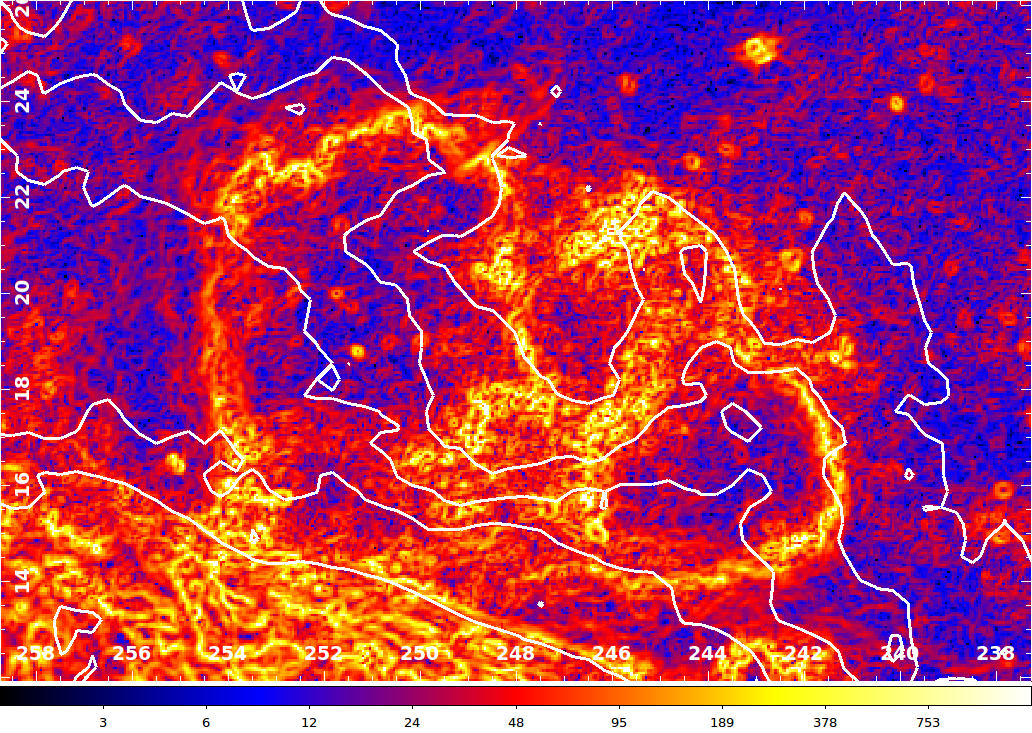}}
\caption{\label{f:ABC_features} \textit{The $|\nabla \textbf{P}|/|\textbf{P}|$ morphologies of the $|\nabla RM|$ features labelled A,B, and C in Fig.~\ref{f:gradPInorm_map} are shown together with the contours of the H$\alpha$ intensity from the all-sky map of \citet{Finkbeiner03}. The units of the H$\alpha$ contours for feature A (top panel), B (middle panel) and C (bottom panel) are 0.1, 2, 3, 4, 16 and 3, 4, 7, 16, 50 and 3, 5, 8, 16, 32 Rayleigh, respectively. Colour scale is the same of Fig.~\ref{f:gradPInorm_cases}.}}
\end{figure}

We note the presence in the $|\nabla \textbf{P}|/|\textbf{P}|$ map (see labels A,B,C in Fig.~\ref{f:gradPInorm_map}) of three structures with no corresponding radio continuum emission. An elongated bubble-like feature (labelled A) is depicted at $(l,b)\approx(6^{\circ},-29^{\circ})$ with a linear size of about $2^{\circ}$. $|\nabla \textbf{P}|/|\textbf{P}|$ and H$\alpha$ intensity correlate, but the H$\alpha$ intensity peaks at $(l,b)\approx(8^{\circ},-30^{\circ})$. A high normalized polarization gradient filament is seen across the feature and displays a single-jump morphology as shown in Fig.~\ref{f:ABC_features}.

A long ($\sim8.4$~deg) and thick ($\sim0.9$~deg) linear structure (labelled B) extends from $(l,b)\approx(346^{\circ}.6,-14^{\circ}.8)$ to $(l,b)\approx(339^{\circ}.3,-17^{\circ}.1)$, and is prominent in both the $|\nabla \textbf{P}|/|\textbf{P}|$ and H$\alpha$ maps, as displayed in Fig.~\ref{f:ABC_features}. The morphology of the $|\nabla \textbf{P}|/|\textbf{P}|$ intensity may consist of two-filaments, both showing a single-jump morphology and slightly diverging towards $(l,b)\approx(339^{\circ}.3,-17^{\circ}.1)$, or a single filament displaying a double-jump morphology. The comparison with the simulated maps favours the two filaments case. The mean width of the upper and lower filaments is about 14\arcmin and 21\arcmin, respectively. The $|\nabla \textbf{P}|/|\textbf{P}|$ intensity correlates with H$\alpha$, which shows a peak of $\sim12$~Rayleigh (1 Rayleigh $= 10^6/4\pi$~photons~cm$^{-2}$~s$^{-1}$~sr$^{-1}$) at $(l,b)\approx(343^{\circ}.0,-16^{\circ}.0)$. The H$\alpha$ contrast of the structure with respect the background is $\sim2$ around the peak position, decreasing to $\sim1.3$ over the rest of the feature. The mean H$\alpha$ contrast corresponds to an emission measure (EM) of about 4.4~pc~cm$^{-6}$. The polarization horizon of the $|\textbf{P}|$ map places an upper limit of about 3~kpc on the distance, implying a length of $\lesssim440$~pc, a width of $\lesssim50$~pc, and a height of $\lesssim900$~pc above the Galactic plane. 
Such extended structures are observed in numerical simulations with sub-Alfv\'enic magnetic field strengths \citep[see][]{Burkhart12}. Linear features may be organized by large-scale magnetic fields and observed perpendicular to the field orientation.


Finally, two prominent shell-like features are seen in $|\nabla \textbf{P}|/|\textbf{P}|$ towards the third Galactic quadrant. The loops are not symmetric, are characterized by high normalized polarization gradients and have centres at $(l,b)\approx(251^{\circ},20^{\circ})$ and $(l,b)\approx(242^{\circ}.5,17^{\circ}.5)$, with radii of about $3^{\circ}$ and $2^{\circ}$, respectively. Along with the extended filaments tracing the edges, a mottled pattern of high $|\nabla \textbf{P}|/|\textbf{P}|$ intensity is also seen on small ($\lesssim 0.5^{\circ}$) scales. No correlation with H$\alpha$ intensity is found for these structures or their surroundings, thus disfavouring the presence of a relevant enhancement of the electron density due to an, e.g., \ion{H}{ii} region. Moreover, the $|\nabla \textbf{P}|/|\textbf{P}|$ filaments are characterized by a single-jump morphological pattern, suggesting the presence of a cusp or shear in the medium.

\section{ISM turbulent regimes}
\label{s:regimes_turbulence}

Polarization gradients are useful means to study turbulence and to constrain the Mach numbers of diffuse and ionized gas because they are sensitive to variations in the free electron density and magnetic field strength. Variations in both the free electron density and magnetic field strength can be induced by strong shocks, thus polarization gradients are effective tracers of transonic and supersonic type turbulence. In addition, turbulence naturally induces discontinuities even in the incompressible limit, so polarization gradients can also indicate subsonic regimes.  

The statistical determination of the sonic Mach number is possible because of the relationship between turbulence regimes and the probability distribution functions (PDFs) of the image intensity distribution \citep{Burkhart12}. The probability distribution function of an image distribution gives the frequency of occurrence of intensities values and can be described by the main statistical moments: mean, variance, skewness, and kurtosis. The higher order moments (skewness and kurtosis) measure the degree of deviation of the PDF from a Gaussian, and are sensitive to the imprints of turbulence encoded in the image intensity distribution. For this reason, their use constitutes a common diagnostic for astrophysical turbulence.

\begin{figure}
\resizebox{9cm}{!}{\includegraphics[angle=0]{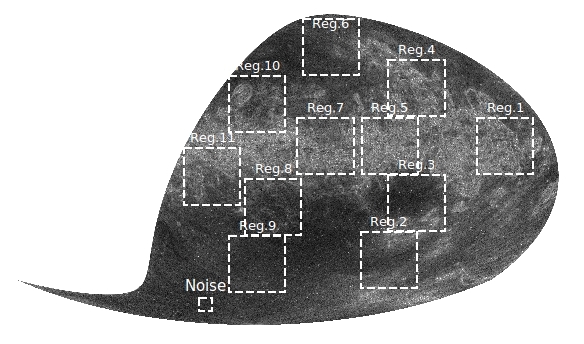}}
\caption{\label{f:fields_map} \textit{The selected boxed regions of the $|\nabla \textbf{P}|/|\textbf{P}|$ map for the statistical determination of the sonic Mach number.}}
\end{figure}

The first-and second-order statistical moments (mean and variance) used here are defined as follows: $\mu_{\xi}=\frac{1}{N}\sum_{i=1}^N {\left( \xi_{i}\right)}$ and $\nu_{\xi}= \frac{1}{N-1} \sum_{i=1}^N {\left( \xi_{i} - \overline{\xi}\right)}^2$, respectively, with \textit{N} the total number of elements and $\xi_{i}$ the distribution of intensities. The third and fourth order moments, skewness, and kurtosis, respectively, are defined as

\begin{equation}
\gamma_{\xi} = \frac{1}{N} \sum_{i=1}^N{ \left( \frac{\xi_{i} - \overline{\xi}}{\sigma_{\xi}} \right)}^3 \quad,
\label{eq:skew}
\end{equation}
\begin{equation}
\beta_{\xi}=\frac{1}{N}\sum_{i=1}^N \left(\frac{\xi_{i}-\overline{\xi}}{\sigma_{\xi}}\right)^{4}-3 \quad,
\label{eq:kurt}
\end{equation}

where $\sigma_{\xi} = \sqrt{\nu_{\xi}}$ is the standard deviation. In this section, we investigate whether there is any large-scale spatial dependence of the ISM turbulence regime as seen in the $|\nabla \textbf{P}|/|\textbf{P}|$ map. To search for spatial variations of the interstellar sonic Mach number, we selected eleven $25^{\circ}\times25^{\circ}$ sub-fields in the $|\nabla \textbf{P}|/|\textbf{P}|$ map, as shown in Fig.~\ref{f:fields_map}. These regions cover the relevant extended features seen in the map of spatial gradient of linear polarization, and sample a wide range of Galactic longitude and latitude. A small test region showing no extended emission was also chosen to characterize the noise statistics.

\begin{figure}[htbp]
\resizebox{9cm}{!}{\includegraphics[angle=0]{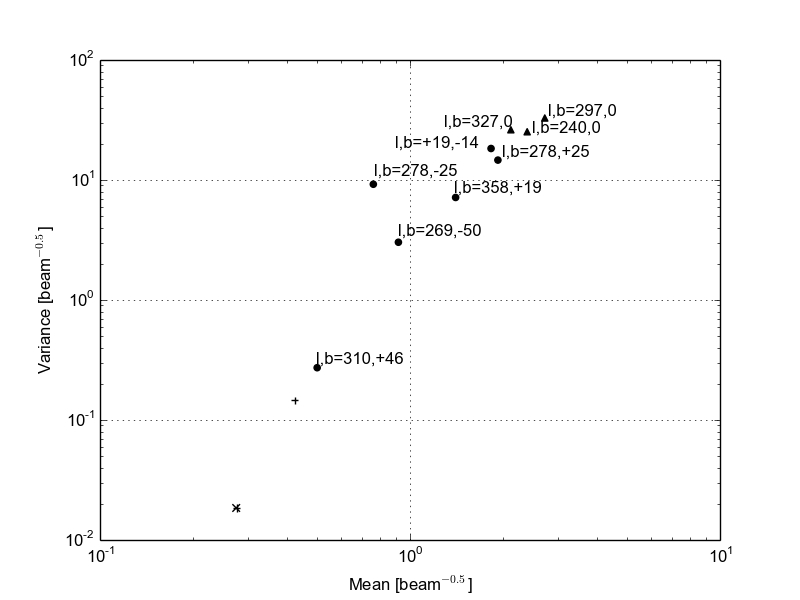}}
\resizebox{9cm}{!}{\includegraphics[angle=0]{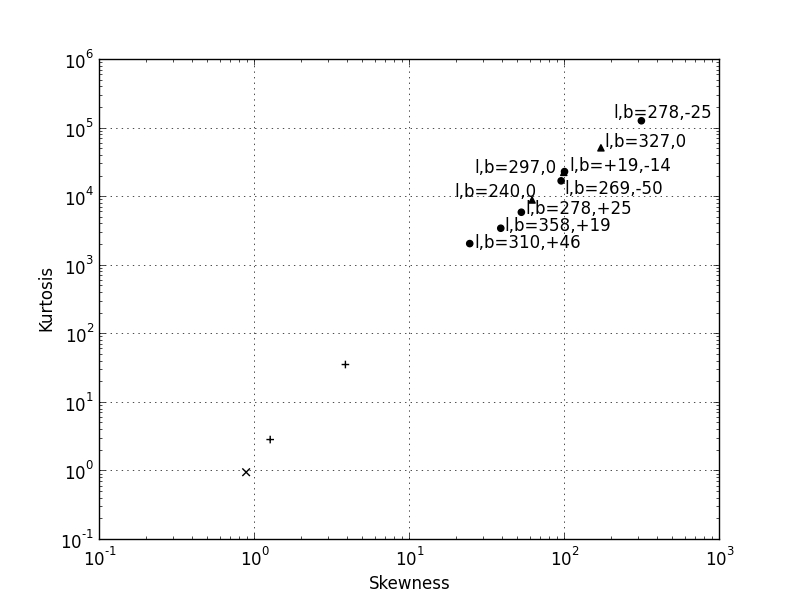}}
\caption{\label{f:momcorr_gradPI} \textit{The mean vs. variance (top) and skewness vs. kurtosis (bottom) for selected regions of the $|\nabla \textbf{P}|/|\textbf{P}|$ map. Noise dominated regions are indicated by crosses, the X symbol referring to the noise test region. Regions close to and off the Galactic plane are marked by filled squares and points, respectively.}} 
\end{figure}

We show correlations between the first-and second-order moments of $|\nabla \textbf{P}|/|\textbf{P}|$ of the twelve selected regions in the top panel of Figure~\ref{f:momcorr_gradPI}, and the correlation between the third-and fourth-order moments in the bottom panel of Figure~\ref{f:momcorr_gradPI}. Tight correlations exist for the moments of all twelve maps, and interesting trends are seen with respect to Galactic latitude and level of noise. In general the noise maps have very low values for all four moments, and the low values of the higher order moments indicate that the noise is Gaussian in nature. Regions with more structure exhibit higher values of the four moments, hence departures from Gaussianity.
In all fields towards the plane, the values of the moments are very high, while the off-plane regions show a range of moments, which may indicate systematic variations in the ISM turbulence regime. 


Figure~\ref{f:momcorr_gradPI} shows the moments for the S-PASS $|\nabla \textbf{P}|/|\textbf{P}|$ data. However if we would like to compare the S-PASS moments with ideal MHD simulations, we must keep in mind two limitations of the observational data and simulated observations.
First, the simulations cannot be normalized by the linear polarization map because they contain only ``Faraday thin'' structures (i.e. the emission and Faraday rotation take places in spatially separated regions) and have no intrinsic background polarization fluctuations. Thus as in the case of the ideal synthetic observations used by \citet{Gaensler11}, one must work with the statistics of the $|\nabla \textbf{P}|$ map alone.  
Second, the observed polarization gradient fluctuations can only be interpreted in terms of turbulence once one has confirmed that the polarization signal is due to foreground Faraday rotation. In this scenario, there should be little correlation between the fluctuations of $|\textbf{P}|$ and Stokes~I. In the case where $\textbf{P}$ or $|\nabla \textbf{P}|$ or $|\nabla \textbf{P}|/|\textbf{P}|$ matches the morphology seen in Stokes~I, the polarized emission has a substantial intrinsic component and therefore does not necessarily describe the foreground turbulence along the line of sight.


\begin{table}[ht!]
\centering
\tiny
\caption{\label{t:regions} Main properties of the selected regions and their correlation degree of $|\textbf{P}|$ with Stokes~I (column 3), as derived by the Pearson's coefficient ($\rho_{p}$) which assumes a linear relationship. The degree of correlation between $|\textbf{P}|$ and $|\nabla \textbf{P}|/|\textbf{P}|$ is indicated in columns 4--5 by the Pearson and Spearman's rank  ($\rho_{s}$) coefficients. The noise test field and region 9 correspond to noise-dominated fields indicated in Fig.~\ref{f:momcorr_gradPI} by the black cross and X symbol, respectively. The Faraday thin regions used in Sect.~\ref{s:moments_analysis} for the quantitative comparison with the MHD simulations are also shown.} 
\begin{tabular}{ll|c|cc|cc}
\hline
\hline
Field & ($l,b$) coord. & $|\rho_{p}|$ & $|\rho_{p}|$ & $|\rho_{s}|$ & Notes \\
 & & $|\textbf{P}|$--I & \multicolumn{2}{ |c| }{$|\textbf{P}|$--$|\nabla\textbf{P}|/|\textbf{P}|$} & & \\
\hline \\
Reg.1 & $240^{\circ},0^{\circ}$ & $0.09$ & $0.31$ & $0.63$ & \\ 
Reg.2 & $269^{\circ},-50^{\circ}$ & $0.35$ & $0.20$ & $0.69$ & \\ 
Reg.3 & $278^{\circ},-25^{\circ}$ & $0.12$ & $0.18$ & $0.70$ & \\ 
Reg.4 & $278^{\circ},+25^{\circ}$ & $0.12$ & $0.29$ & $0.61$ & \\ 
Reg.5 & $297^{\circ},0^{\circ}$ & $0.46$ & $0.15$ & $0.48$ & Faraday thin \\ 
Reg.6 & $310^{\circ},+46^{\circ}$ & $0.37$ & $0.45$ & $0.60$ & \\ 
Reg.7 & $327^{\circ},0^{\circ}$ & $0.16$ & $0.21$ & $0.49$ & Faraday thin \\ 
Reg.8 & $350^{\circ},-28^{\circ}$ & $0.06$ & $0.18$ & $0.11$ & Faraday thin \\ 
Reg.9 & $357^{\circ},-56^{\circ}$ & $0.52$ & $0.36$ & $0.37$ & Noise- dominated \\ 
Reg.10 & $358^{\circ},+19^{\circ}$ & $0.19$ & $0.23$ & $0.47$ & Faraday thin \\ 
Reg.11 & $19^{\circ},-14^{\circ}$ & $0.09$ & $0.17$ & $0.26$ & Faraday thin \\ 
Noise & $74^{\circ},-76^{\circ}$ & $0.23$ & $0.18$ & $0.17$ & Noise- dominated \\ 
\hline \\
\end{tabular} 
\end{table}


We show the Pearson correlation coefficients of $|\textbf{P}|$ with Stokes~I in Table~\ref{t:regions} for each region. The Pearson coefficient measures the linearity of a relationship, while the Spearman's rank coefficient measures any correlation. We tested it because the linear relationship between Stokes~I and $|\textbf{P}|$ is expected in the Faraday thin approximation. We see that in general strong correlations do not exist between $|\textbf{P}|$ with Stokes~I in these maps. 

Since there is no strong correlation between $|\textbf{P}|$ and Stokes~I, we can assume that the bulk of the emission and features seen in $|\nabla \textbf{P}|/|\textbf{P}|$ arise from (foreground) Faraday fluctuations in the turbulent MIM. Moreover, for cases that are properly described by the Faraday thin approximation, $|\nabla \textbf{P}|/|\textbf{P}|$ will tend to $|\nabla \textbf{P}|$ and the quantitative comparison with simulations is meaningful. Thus we can make a comparison between the statistics of $|\nabla \textbf{P}|$ of the S-PASS observations and the MHD turbulence simulations of \citet{Burkhart12}, since the emission in both arises from turbulent fluctuations of $n_e$ and $B$ along the line of sight. This is the aim of the next section.


\subsection{Moments of $|\nabla \textbf{P}|$ and the sonic Mach number}
\label{s:moments_analysis}

To test the goodness of the ``Faraday thin'' approximation for the selected regions, we look at the correlation degree between $|\textbf{P}|$ and $|\nabla \textbf{P}|/|\textbf{P}|$. To this aim we use both the Pearson and the Spearman rank ($\rho_{s}$) correlation coefficients, with $\rho_{s}$ defined as
\begin{equation}
\rho_{s} = 1 - \frac{6\sum d_{i}^{2}}{N(N^{2}-1)} \quad,
\end{equation}
with $d_{i}$ the difference in paired ranks of $|\textbf{P}|$ and $|\nabla \textbf{P}|/|\textbf{P}|$ intensities. As discussed in Sect.~\ref{s:pol_grad}, a negligible correlation is expected between $|\textbf{P}|$ and $|\nabla \textbf{P}|/|\textbf{P}|$ in the Faraday thin regime with no intrinsic background polarization fluctuations. However, intrinsic background polarization fluctuations affect our data, so we assume $|\rho|<0.5$ as a threshold for low or mild correlation. 




The relationship between Pearson and Spearman's rank correlation coefficients of $|\textbf{P}|$ and $|\nabla \textbf{P}|/|\textbf{P}|$ for each region is displayed in  Table~\ref{t:regions}. 
In general, $\rho_{p}$  and $\rho_{s}$ point to the same results only in the following (extreme) cases: a linear relationship exists between the inspected variables, or the variables are uncorrelated. A sample of seven regions is selected: two are noise-dominated and two or three fields lie towards to/off the Galactic plane, respectively. The noise (dominated) regions display the lowest degree of correlation for both the correlation coefficients. For each selected region, we check the degree of correlation between $|\nabla \textbf{P}|$ and $|\nabla \textbf{P}|/|\textbf{P}|$ maps, finding mild or high values (i.e. $>0.5$) of the Spearman's rank correlation coefficient. 
We then assume the Faraday thin approximation to hold for these few regions that sample a wide range of Galactic longitude and latitude. The small test region showing no extended emission is included to characterize the noise statistics. To search for spatial variations of the interstellar sonic and number, results from the statistical analysis of signal distributions of boxed $25^{\circ}\times25^{\circ}$ sized sub-fields in the $|\nabla \textbf{P}|$ map are then matched with numerical simulations of isothermal compressible MHD turbulence, since the emission in both arises from turbulent fluctuations of $n_e$ and $B$ along the line of sight.

Following \citet{Burkhart12}, we performed a statistical determination of the sonic Mach number by using the probability distribution function moments of the image distribution. 
In the framework of polarization gradients, the relationship between the moments of the $|\nabla \textbf{P}|$ map and the sonic Mach number has been studied by \citet{Burkhart12}. As a consequence of the shock fronts creating more discontinuities and sharper gradients, a systematic increase in all four moments with increasing sonic Mach number is found. 

In the analysis that follows, we employ the same set of MHD simulations that was used in the study by \citet{Burkhart12} to estimate the sonic Mach number in S-PASS data. The database of 3D numerical simulations of isothermal compressible (MHD) turbulence is generated by using the MHD code of \citet{ChoLazarian03} and varying the input  values for the sonic and Alfv\'{e}nic Mach number. We scale the simulations to physical units, adopting typical parameters for warm ionized gas. We assume an average density of 0.1~cm$^{-3}$ \citep{Gaensler11}. The simulations are assumed to be fully ionized, and we do not include the effects of partial ionization. To make the maps of $|\nabla \textbf{P}|$, we first calculate the line of sight rotation measure at each pixel, then take the gradient of this rotation measure map and convert it to $|\nabla \textbf{P}|$. For more information and details on these simulations, we refer the reader to \citet{ChoLazarian03} and \citet{Burkhart12}.

An additional effect that must be considered when applying the moments to estimate the sonic Mach number in $|\nabla \textbf{P}|$ data is the question of the telescope resolution. It is clear from the analysis of \citet{Burkhart12} that smoothing changes the distribution of maps of $|\nabla \textbf{P}|$. As the resolution of maps of Stokes~Q and U decreases, so do the moments. Thus one needs to take the smoothing of the data into account when comparing PDFs of $|\nabla \textbf{P}|$. We address this issue in the current study by applying four-pixel smoothing (equivalent to 10.75\arcmin smoothing assuming that a scaling of one of our pixels at 512x512 resolution is 0.048 degrees) to the simulated $|\nabla \textbf{P}|$ maps. This conversion is appropriate because the actual beam size of interest is a combination of the size scale of the beam in relation to the size scale of the turbulent injection. Given that the S-PASS maps we investigate are 25x25 degrees wide with a polarization horizon of $\approx 3$~kpc, this means we are sampling emission in these boxes up to scales of $\approx 1$~kpc. This is larger than the scale of the injection of turbulence in the Galaxy, but compatible with the large-scale driving of turbulence in our simulations. 


\begin{table*}[htbp]
\centering
\caption{\label{t:nonorm_regions} Summary of high-order moments analysis obtained from selected ``Faraday thin'' regions. Noise and Region 9 correspond to noise-dominated fields indicated by the shaded box in Fig.~\ref{f:momcorr_nonorm_gradPI}.}
\begin{tabular}{llcccccc}
\hline
\hline \\
Field & ($l,b$) coordinates&Mean&Variance&Skewness & Kurtosis & $M_{s}$ regime\\ 
 & &mJy~beam$^{-1.5}$&[mJy~beam$^{-1.5}]^{2}$& & & \\ 
\hline \\
Reg.5 & $297^{\circ},0^{\circ}$ & $0.058$ & $0.041$ & $1.70$ & $4.60$ & SUB/TRANS \\ 
Reg.7 & $327^{\circ},0^{\circ}$ & $0.16$ & $0.11$ & $1.56$ & $3.42$ & SUB/TRANS \\ 
Reg.8 & $350^{\circ},-28^{\circ}$ & $0.041$ & $0.030$ & $2.26$ & $9.09$ & TRANS \\ 
Reg.9 (Noise) & $357^{\circ},-56^{\circ}$ & $0.015$ & $0.0062$ & $0.76$ & $1.18$ & \\ 
Reg.10 & $358^{\circ},+19^{\circ}$ & $0.099$ & $0.081$ & $2.13$ & $6.90$ & TRANS \\ 
Reg.11 & $19^{\circ},-14^{\circ}$ & $0.088$ & $0.093$ & $3.03$ & $19.65$ & TRANS/SUPER \\ 
Noise & $74^{\circ},-76^{\circ}$ & $0.014$ & $0.0062$ & $0.75$ & $0.75$ & \\ 
\hline \\
\end{tabular} 
\end{table*}


\subsubsection{Moments analysis: results}

We calculated the values of the first four statistical moments of the sub-field regions of the S-PASS selected Faraday thin regions and compared the values with what is found in MHD simulations for different values of the sonic Mach number \citep[see][]{Burkhart12}. We used the full 512x512 sub-fields for calculating of the skewness and kurtosis in order to obtain meaningful estimates \citep{DudokdeWit04}. However, simulations show that this condition might be relaxed somewhat, while still retaining valid results \citep{Burkhart10}. Table~\ref{t:nonorm_regions} lists the results of calculating the moments for the S-PASS regions. Columns 3--6 show the all-map values of the skewness and kurtosis. We compare these values with the findings of \citet[][]{Burkhart12}, who report values of skewness and kurtosis for sonic Mach numbers ranging from subsonic to supersonic values of upwards of 10. Subsonic cases have typical values of skewness $\lesssim 1.6$ and values of kurtosis $\lesssim 5$. Transonic simulations show values of skewness from 1.6--3 and kurtosis values from 5--17. Values higher than these are generally in the range of supersonic turbulence.

\begin{figure}
\resizebox{9cm}{!}{\includegraphics[angle=0]{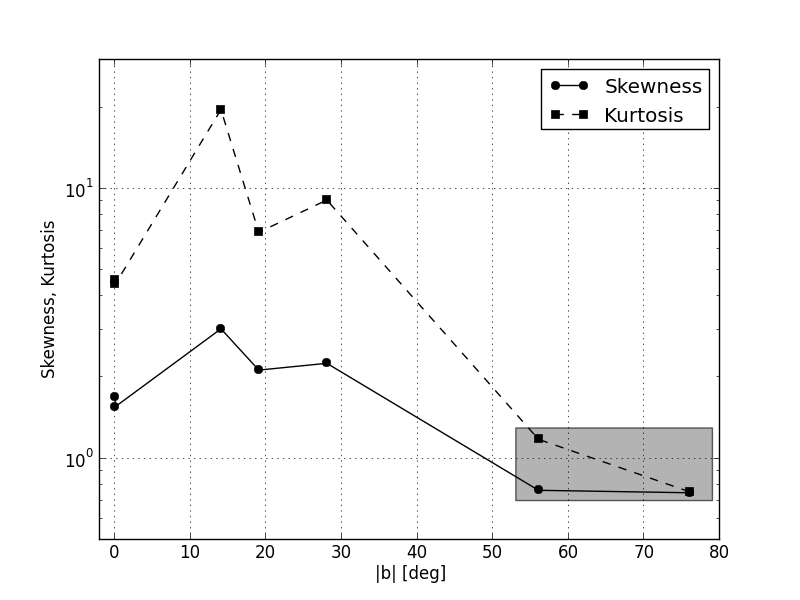}}
\caption{\label{f:momcorr_nonorm_gradPI} \textit{The skewness and kurtosis as a function of Galactic latitude for the ``Faraday thin'' selected regions of the $|\nabla \textbf{P}|$ map. Noise dominated regions are indicated by the shaded box.}}
\end{figure}

From the values presented in Table~\ref{t:nonorm_regions}, a tight correlation can be found between mean and variance and between skewness and kurtosis, with the highest sonic Mach numbers corresponding to the highest values of moments and the lowest corresponding to low Mach numbers. We find this correlation behaviour of the four main moments also in the simulations, regardless of the addition of beam smoothing. Noise-dominated regions show systematically low values of the mean, the variance, and to a lesser extent, the skewness and kurtosis. In order to highlight any spatial dependence of the higher order moments for the selected regions in our data, we plot the calculated all-map skewness and kurtosis as a function of Galactic coordinates in Fig.~\ref{f:momcorr_nonorm_gradPI}. A clear correlation of both skewness and kurtosis as a function of Galactic latitude is seen, with the mid Galactic latitude regions showing slightly higher skewness and kurtosis. However, this trend implies only a weak variation in the sonic Mach number. As a consequence, no systematic clustering of the sonic Mach number values as a function of Galactic latitude is found, as is seen in the lower panel of Fig.~\ref{f:momcorr_nonorm_gradPI}.


\section{Discussion}
\label{s:discussion}

Figure~\ref{f:Gb_counts_gradpi} shows that the distribution peak position of the $|\nabla \textbf{P}|/|\textbf{P}|$ intensities at Galactic latitudes $|b|<15^{\circ}$ differs systematically from the peak positions at higher latitude ranges. The evident shift towards lower intensity values for higher Galactic latitudes observed towards the inner regions relates to the vertical extension of the Carina-Sagittarius spiral arm. Assuming a distance of $\sim2$~kpc to this arm, the corresponding height above the Galactic disk is $\lesssim500$~pc, in agreement with the typical vertical extension of other arms \citep{Feitzinger86} and the scale height of the arms in the \citet{Taylor93} model of the free electron density.

Moreover, the $|\nabla \textbf{P}|/|\textbf{P}|$ intensity shows a dependence on both Galactic longitude and latitude. In Figure~\ref{f:Gb_counts_gradpi} we find that $|\nabla \textbf{P}|/|\textbf{P}|$ emission is roughly approximated by a log-normal distribution, with deviations towards small $|\nabla \textbf{P}|/|\textbf{P}|$ fluctuations. A log-normal distribution is expected for pure density perturbations in a 3D turbulent and isothermal flow \citep[see e.g.][]{Vazquez-Semadeni94} and is observed for density tracers in the WIM such as the emission measure \citep[][]{Hill08}. However, our observations trace an additional part of the MIM, called the warm partially ionized medium (WPIM, see \citep[][]{Heiles01,HeilesHaverkorn12}). This component is colder than the WIM ($T\approx 5000$~K) and does not emit H$\alpha$ emission. However, it does contribute to the Faraday rotation in the medium. Therefore, our study traces both the WIM and WPIM, as opposed to the \citet{Hill08} results that are only sensitive to the WIM. Furthermore the presence of magnetic fields in the MIM also influences the PDF of the perturbations \citep{Kowal07,Molina12,BurkhartLazarian12b}. Therefore, the deviations from a log-normal distribution we point out in Section~\ref{s:variations} may be explained by the sensitivity of the $|\nabla \textbf{P}|/|\textbf{P}|$ emission to both density and magnetic fluctuations. In addition, the presence in our data of a polarization horizon at a distance of about 3~kpc also affects the observed PDF of $|\nabla \textbf{P}|/|\textbf{P}|$. As a consequence, most of the diffuse polarized emission seen in the $|\nabla \textbf{P}|/|\textbf{P}|$ map in Figure~\ref{f:gradPInorm_map} is generated within the nearest spiral arm. The short path length limits the number of the observed independent fluctuations, thus affecting the convergence to a Gaussian.

In Figure~\ref{f:momcorr_nonorm_gradPI} small variations of the sonic Mach number are observed in the Faraday-thin selected fields towards and out of the Galactic plane at medium latitudes. These variations may be consistent with turbulence in the MIM being driven by supernova explosions \citep[see simulations by][]{Hill12}, but they may also be explained by the presence of spatially extended and nearby objects triggering turbulence in (and interacting with) the surrounding medium. These features affect the statistical inference of the turbulent regime in some of the target regions. Since it is not possible to separate their contribution from the bulk of the $|\nabla \textbf{P}|$ features, they were not masked.

Finally, we report the detection of a prominent spur-like feature (see Sect.~\ref{s:imaging}) clearly seen towards the edge of the third Galactic quadrant in the $|\nabla \textbf{P}|/|\textbf{P}|$ map but not traced by H$\alpha$ intensity. Because of the high latitude, extinction is not likely to be responsible for the H$\alpha$ non-detection, thus disfavouring a thermal electron density enhancement. Moreover, this magnetic structure has no radio continuum counterpart, suggesting that the direction of the magnetic field is mainly pointed towards the observer. The large angular size of this magnetic feature, which is specific of to the $|\nabla \textbf{P}|/|\textbf{P}|$ map, suggests the structure is nearby. These characteristics, along with its shape, support its association with Loop~I, a local feature of the ISM \citep{Berkhuijsen71,Heiles79} associated with an expanding SNR \citep{Spoelstra72,Heiles98,Wolleben07}.

\section{Summary and conclusions}
\label{s:conclusion}

Normalized spatial gradients of the polarization vectors have been used for the first time to map the entire southern sky. The large sky coverage allows the exploration of cases not treated by the previous studies of \citet{Gaensler11} and \citet{Burkhart12}. The S-PASS $|\nabla \textbf{P}|/|\textbf{P}|$ map displays a wealth of filamentary structures with typical widths down to the angular resolution. The emission is characterized by a polarization horizon of about 3~kpc, implying density and magnetic fluctuations down to a linear scale $<10$~pc given the angular size of the S-PASS beam. An extended and patchy pattern of $|\nabla \textbf{P}|/|\textbf{P}|$ intensity is found within the third and fourth Galactic quadrants at high ($b\lesssim -60^{\circ}$) latitudes towards the south Galactic pole.

Two different morphologies (i.e. ``single'' or ``double'' jump profiles) corresponding to different MHD turbulence cases (i.e. low or high sonic Mach numbers) are observed, thus supporting the predictions of numerical simulations \citep{Burkhart12}. 
Normalized spatial gradients of the polarization vector are effective tracers of extended and Faraday rotating features, such as \ion{H}{ii} regions and evolved SNRs. Indeed we clearly recognize the two known nearby and old SNRs Antlia and G\,353-34. In addition, by combining the information from both the $|\nabla \textbf{P}|/|\textbf{P}|$ and H$\alpha$ intensity maps we can highlight the presence of both electron density and magnetic structures, in agreement with simulations \citep{Burkhart12}.

Although multiple scales of energy injection are expected in the ISM \citep{NotaKatgert10}, instabilities triggered by supernova events and Galactic shear in the ISM are expected to mainly generate and sustain interstellar MHD turbulence \citep[][]{MacLow04,Hill12}. Observational studies of turbulence in the warm and ionized ISM also indicate a spectral index matching that of the \citet{Goldreich&Sridhar95} theory of Alfv\'{e}nic turbulence, consistent with a weakly compressible medium. This is the case for transonic turbulence as shown by \citet{Hill08}, who estimated the sonic Mach number by comparing statistics of H$\alpha$ WHAM data with simulations. By applying a moment analysis to a number of fields, we extend it to the MIM and confirm the earlier result of these authors, finding lines of sight to be consistent with $M_{s} \lesssim 2$ (see Table~\ref{t:nonorm_regions}). 

The use of the spatial gradient of linear polarizations combined with a robust statistical analysis makes mapping of the sonic and Alfv\'{e}nic Mach numbers spatial variations in the MIM a feasible and mandatory aim of forthcoming radio observations at high angular resolution. These studies will allow us to gain complementary insight into the turbulence and shocks in the ionized ISM over a wide range of plasma $\beta$-parameter regimes. To gain a complete picture of Mach numbers and spatial variations in the MIM, complete sky coverage is needed, requiring a corresponding high resolution and sensitivity survey of the northern sky.

\begin{acknowledgements}
The authors thank the referee, Steven Spangler, for providing detailed comments and helpful suggestions in the preparation of the final manuscript.
This work has been carried out in the framework of the S-band Polarisation All Sky Survey collaboration (S-PASS). The Parkes Radio Telescope is part of the Australia Telescope National Facility, which is funded by the Commonwealth of Australia for operation as a National Facility managed by CSIRO. The research leading to these results has received funding from the European Union's Seventh Framework Programme (FP7/2007-2013) under grant agreement number 239490. This work is part of the research programme 639.042.915, which is (partly) financed by the Netherlands Organization for Scientific Research (NWO). Support for B.B. and A.L. comes from the NSF grant AST 1212096, and the Center for Magnetic Self-Organization in Laboratory and Astrophysical Plasmas (CMSO). B.B. acknowledges Vilas Associate Awards and the hospitality of the International Institute of Physics (Natal). Parts of this research were conducted by the Australian Research Council Centre of Excellence for All-sky Astrophysics (CAASTRO), through project number CE110001020. B.M.G. acknowledges the support of the Australian Research Council through an Australian Laureate Fellowship (FL100100114).
\end{acknowledgements}

\end{document}